\documentclass[11pt]{article}
\usepackage{amsmath,amssymb}
\usepackage{graphicx,color}

\def\({\left(}
\def\){\right)}

\newcommand{\be}{\begin{equation}}
\newcommand{\ba}{\begin{eqnarray}}
\newcommand{\ea}{\end{eqnarray}}
\newcommand{\ee}{\end{equation}}

\newcommand{\la}{\langle}
\newcommand{\lb}{\rangle}

\begin{document}

\begin{titlepage}
\thispagestyle{empty}

\begin{flushright}

\end{flushright}

\bigskip

\begin{center}
\noindent{\Large \textbf{Holographic Entanglement Entropy for $d=4$ $\mathcal{N}=2$ SCFTs \\ in F-theory}}\\
\vspace{15mm} Mitsutoshi Fujita\footnote{email: mf29@uw.edu}
\vspace{1cm}

{\it Department of Physics, University of Washington, Seattle, WA 98195-1560, USA} \\

\vskip 3em

\begin{abstract}{We compute the holographic entanglement entropy in the  gravity with higher curvature terms dual to $d=4$ $\mathcal{N}=2$ SCFTs in F-theory using the method proposed in arXiv:1011.5819. The log term of this entanglement entropy reproduces the $A$-type anomaly of the dual SCFTs in F-theory using the gravity dual with the higher derivative term, which, using the field redefinition, can be transformed to the Gauss-Bonnet term in the subleading order of a derivative expansion. Our analysis shows that the $1/N^2$ correction for the central charges should be produced from the gravity including higher derivative terms up to curvature squared terms. To obtain consistent results, we discuss the holographic c-theorem.} 
\end{abstract}
\end{center}
\end{titlepage}

\section{Introduction}
The entanglement entropy is an important quantity for quantum field theories. The entanglement entropy is non-vanishing even at zero temperature and it is proportional to the degrees of freedom. Moreover,  it can classify the topological phases described by the ground state of the FQHE or the Chern-Simons effective theory~\cite{wenx,kita,DFLN}. In the context of the AdS/CFT correspondence~\cite{juan,anom}, the formula of holographic Entanglement entropy (holographic EE)~\cite{RT} for CFTs has been proposed as the area of minimal surface $\gamma_A$ which ends on the boundary of the region $A$ as follows:
\ba
S_A=\dfrac{\text{Area}(\gamma_A)}{4G_N}, \label{HOL1}
\ea 
where $G_N$ is the Newton constant in the Einstein gravity. Here, $\gamma_A$ is required to be homologous to $A$. It has been shown~\cite{RT,Sol,LNST,CaHu,Solog} that the log term of the entanglement entropy agrees with the computation in the gravity side. In~\cite{FRWT}, the topological entanglement entropy is clarified via the AdS/CFT correspondence.
 Moreover, it is shown~\cite{HHM} that using holographic EE,  the correlation in the holographic system arises from the quantum entanglement rather than the classical correlation. The review of the holographic EE are given in~\cite{NRT} and in the review part of~\cite{Tonni:2010pv}. 
 
While the holographic EE of~\cite{RT} can be used for the Einstein gravity with no higher derivative terms, it is shown~\cite{Myers:2010xs,Myers:2010tj,Casini:2011kv,Hung:2011ta,Hung:2011nu} that for the spherical surface of the boundary of $A$, the proposal for holographic EE \eqref{HOL1} can be extended to certain higher curvature theories such as Gauss-Bonnet gravities. 
It is proposed that using the conformal mapping, EE across the spherical surface $S^2$ with the radius $L$ can be related with the thermal entropy on $\mathbb{R}\times H^{3}$ with $T=1/(2\pi L)$ and this thermal entropy is given by the black hole entropy in the gravity side. It is also shown that the log term of the holographic EE is then proportional to the $A$-type anomaly.

Thus, it is considered to be interesting to apply these results for higher derivative corrections in the holographic EE in some string models.
An interesting model is the 5-dimensional curvature gravity instead of beginning with the 10-dimensional theory, where the curvature-squared terms arise from the D-branes~\cite{bng,ode,Buchel:2008vz,kp}. Here, the higher curvature terms correct also the cosmological constant which should be determined to reproduce the central charges $a$ and $c$ of the 4-dimensional theory. Using the holographic Weyl anomaly~\cite{bng,ode,holo4}, these terms affect the two central charges $a$ and $c$ in the subleading order of $N$. Moreover, this curvature-squared terms  appear in the context of perturbative expansion of the string theory and it is known that they can be modified by field redefinitions. 

A purpose of this paper is to compute the holographic EE and $A$-type anomaly in the gravity dual to $d=4$ $\mathcal{N}=2$ SCFTs in F-theory. Here, this gravity dual is described by the curvature gravity theory of above the string model. 
These SCFTs include the theory with $G=D_4,E_n,...$ flavor symmetries. Except for $D_4$ all of them are isolated theories, i.e. their gauge couplings are frozen and a perturbative description is not available.  There are previous works for holographic computation of the central charges in these SCFTs~\cite{anom2,at} in which they analyzed Chern-Simons terms in the dual string theory. Their result implies that the AdS/CFT correspondence can be applied even for the finite string coupling. On the other hand, the curvature term of our gravity theory makes the analysis difficult since for $d=5$ these terms depend on the metric of the spacetime. According to~\cite{Buchel:2008vz}, we choose the cosmological constant to reproduce the central charges $a,c$ of the dual SCFT in the context of the holographic Weyl anomaly. As stated above, we can also transform the effective $d=5$ curvature gravity obtained from the F-theory compactifications to the $d=5$ Gauss-Bonnet gravity in the subleading order of $l_s^2/L^2$ expansion using the field redefinition, which is important in the context of the holographic c-theorem.
 And we derive the holographic EE computing the black hole entropy.  

This paper is organized as follows: In section 2, we briefly review the Wald's entropy formula which is important for computing the black hole entropy in the presence of the higher curvature terms. In section 3, we review the construction of the Type IIB background after including the backreaction of the $N$ D3-branes at singularities in F-theory. 
In the beginning of section 4, we review the construction of the effective curvature gravity from F-theory. In subsection 4.1, we analyze what terms contribute to the central charges in SCFTs.  In subsection 4.2, by using the holographic Weyl anomaly, 
we determine the parameters of the gravity to realize the central charges in the field theory side. In subsection 4.3, we generalize the argument in section 4 for the general SCFTs in F-theory.
 In section 5, we compute the holographic EE and the $A$-type anomaly using the $d=5$ effective curvature gravity. We also consider the transformation of the effective $d=5$ curvature gravity to the $d=5$ Gauss-Bonnet gravity using the field redefinition.

\section{Holographic EE as the Wald's entropy formula}
In this section, we briefly review the holographic EE formula in the presence of the higher curvature terms.
It was conjectured~\cite{Myers:2010xs,Myers:2010tj} that EE across the sphere $S^2$ with the radius $L$ can be given by the black hole entropy in the gravity side and the contribution to the EE is then proportional to the coefficient of $A$-type anomaly. This is since when we write the pure AdS by the topological BH with the horizon of radius $L$, there appear two hyperbolic spaces inside and outside the horizon, respectively. By the  conformal mapping, the thermal entropy with temperature $T=1/(2\pi L)$ in a hyperbolic space is shown to be equal to the EE across the sphere $S^2$ as follows:
\ba
S_{EE}=S_{thermal}.
\ea
Thus, the horizon entropy of the topological BH should be considered as EE of the dual theory using the AdS/CFT correspondence. 
Moreover, two hyperbolic spaces are considered as two separated space when we compute the entanglement entropy. 

We start with the general gravity theory in $AdS$ with higher derivative terms as follows: 
\ba
S=\dfrac{1}{2\kappa^2_5}\int d^5x\sqrt{-g}F(R^{\mu\nu}{}_{\rho\lambda})+S_{matter}, \label{ACT03}
\ea
where $F(R^{\mu\nu}{}_{\rho\lambda})$ is constructed from the Riemann tensors and $\mu,\nu,\rho,\lambda=0,1,2,3,4$. 
The Einstein equation for \eqref{ACT03} is given by
\ba
F'_{(\mu}{}^{\rho\lambda\alpha}R_{\nu)\rho\lambda\alpha}-\dfrac{F(R^{\mu\nu}{}_{\rho\lambda})}{2}g_{\mu\nu}+2\nabla^{\rho}\nabla^{\lambda}F'_{\mu\rho\nu\lambda}=\kappa^2_5T_{\mu\nu},
\ea
where $F'_{\mu\nu\rho\lambda}=\delta F(R^{\alpha\beta}{}_{\gamma\delta})/\delta R^{\mu\nu\rho\lambda}$, round brackets describe the symmetrization, and $\nabla_{\mu}$ is the  covariant derivative. Note that the tensor $F'_{\mu\nu\rho\lambda}$ is antisymmetrized and symmetrized as seen in the Riemann tensor. 
To compute the holographic EE, we should choose the $AdS_5$ topological black hole solution with its radius $L$ where the terms including the covariant derivatives vanish since Riemann curvature of maximally symmetric space like $AdS_5$ can be described by the metric (see appedix A):
\be
ds^2=\dfrac{d\rho^2}{\frac{\rho^2}{L^2}-1}-\Big(\frac{\rho^2}{L^2}-1\Big)d\tau^2+\rho^2(du^2+\sinh^2ud\Omega_{2}^2), \label{MET05}
\ee
where $d\Omega_2^2$ represents the metric of $S^2$ and the BH horizon is at $\rho ={L}$.
Then, the holographic EE of the topological black hole is given by extremizing Wald's entropy formula as follows:
\ba
S_{entropy}=-2\pi \int_{horizon}d^3x\sqrt{h}\dfrac{\delta S}{\delta R^{\mu\nu}{}_{\rho\lambda}}\epsilon^{\mu\nu}\epsilon_{\rho\lambda}.
\ea
where $\epsilon^{\mu\nu}$ is the binormal to the horizon satisfying $\epsilon_{\mu\nu}\epsilon^{\mu\nu}=-2$ for \eqref{MET05}.
According to~~\cite{Sinha:2010ai,Myers:2010xs,Myers:2010tj,Casini:2011kv}, 
the integrand can be related with the $A$-type anomaly in the SCFT side as follows:
\ba
a_4=-\pi^2L^3\dfrac{\delta S}{\delta R^{\mu\nu}{}_{\rho\lambda}}\epsilon^{\mu\nu}\epsilon_{\rho\lambda}\Big|_{horizon}. \label{CEN5}
\ea
This $a_4$ is also given by the $a$-function used in the context of the holographic c-theorem if it is evaluated at an AdS vacua.
For the case of Einstein gravity $F(R^{\mu\nu}{}_{\rho\lambda})=R-\Lambda$, in addition, the $A$-type anomaly is given by
\ba
a=\dfrac{\pi^2L^3}{\kappa_5^2}.
\ea
We will use \eqref{CEN5} in section 4 to estimate the corrections 
 for the $A$-type anomaly and in section 5, use for computing the $A$-type anomaly for the $\mathcal{N}=2$ SCFT in F-theory. 
\section{The gravity dual to $d=4$ $\mathcal{N}=2$ SCFTs}

In this section, we review the construction of the Type IIB background dual to  $\mathcal{N}=2$ SCFTs.
It has been known that $\mathcal{N}=2$ SCFTs arise as worldvolume theories of D3-branes moving near 7-branes, namely 3-branes in F-theory~\cite{ft1,ft2,ft3,ft4,ft5}. 
The simplest theory is the worldvolume theory of D3-branes in a $\mathbf{Z}_2$ orientifold and becomes a $USp(2N)$ $\mathcal{N}=2$ gauge theory with an antisymmetric and four fundamental hypermultiplets in the low energy limit. Its gravity dual is the type IIB supergravity on an orientifold of $AdS_5\times S^5$~\cite{Fayyazuddin:1998fb,Aharony:1998xz}.

For the gravity side, we consider the type IIB string theory on $AdS_5\times X_{5}$ with $N$ units of $F_5$ flux on $X_5$ taking the near-horizon limit. 
The metric is as follows:
\be
ds^2=\dfrac{L^2}{z^2}(d\vec{x}^2+dz^2)+L^2ds^2_{X_5}, 
\ee
where $L=(4\pi g_s\Delta N)^{1/4}l_s$ is the radius of $AdS_5$~\cite{Gubser:1998vd} 
satisfying $L^4\text{Vol}(X_5)=L^4_{S^5} \text{Vol}(S^5)$.
Here, $X_{5}$ is the 5-sphere
\be
\{|z_1|^2+|z_2|^2+|z_3|^2=const \}, \label{MET6}
\ee
 with the periodicity $2\pi/\Delta$ along the phase
 $\phi\equiv$arg$(z_3)$ and
the metric of $X_{5}$ is given by
\be
ds^2_{X_5}=d\theta^2+\sin^2\theta d\phi^2 +\cos^2\theta d\Omega^2_3,
\ee
 where $d\Omega ^2_3$ is the metric of $S^3$, $0\le \theta  \le \pi/2$. At the singularity $\theta =0$, the metric becomes $S^3$.  It is known that the singularities are related with flavor groups of 7-branes, which are called $G$-type 7-branes for the $G$-type singularity ($G=H_0,H_1,H_2,D_4,E_6,..$). This 7-brane then wraps whole $AdS_5\times S^3$. Indeed, the transverse direction $Re(z_3),Im(z_3)$ operated by $\mathbb{Z}_2$ orbifold is the transverse direction of an O7-brane for $\Delta=2$. The number of 7-branes is related with the deficit angle $\Delta$ and 
 is given by $n_7=12(1-1/\Delta)$.

The gravity action dimensionally reduced to 5-dimension is given by
\be
I\equiv \int d^{5}x\sqrt{-g}\mathcal{L}=\dfrac{1}{2\kappa_5^2}\int d^5x\sqrt{-g}(R+12/L^2+...), \label{AC09}
\ee
where $2\kappa^2_{10} =(2\pi)^7g_s^2l_s^8$ and we defined 
\be
\dfrac{1}{2\kappa_5^2}\equiv \dfrac{L^5\text{Vol} (X_5)}{2\kappa_{10}^2} \label{DEF10}
\ee Here, dots include the matter fields and the higher derivative corrections. To describe the gravity dual of the $d=4$ $\mathcal{N}=2$ SCFT with flavor, moreover, we should 
include the curvature-squared terms of 7-branes since these terms affect the $A$-type anomaly as the $1/N$ correction (see appendix B).  
In next section, we start with the 5-dimensional effective curvature gravity from F-theory instead of considering Type IIB D-branes.

\section{Effective curvature gravity dual with the orientifold}

It has been known~\cite{Buchel:2008vz,kp} that if the central charges of the dual CFT satisfy
\ba
c\sim a \gg 1 \ \text{and} \ |c-a|/c \ll 1,
\ea
 the effective gravity theory is described by the curvature gravity with the higher curvature term coupled to a negative cosmological constant.
Using the constants $a_1$, $a_2$, and $a_3$, the effective curvature gravity is given by
\ba
S=\dfrac{1}{2\kappa_{5}^2}\int d^5x\sqrt{-g}\Big(R+\dfrac{12}{ \tilde{L}^2}+a_1R^2+a_2R_{\mu\nu}R^{\mu\nu}+a_3R_{\mu\nu\rho\lambda}R^{\mu\nu\rho\lambda} \Big), \label{ACT6}
\ea
where $-12/\tilde{L}$ describes the cosmological constant.
It is known that in the case of $\mathcal{N}=2$ SCFT dual to $AdS_5\times S^5/\mathbb{Z}_2$, moreover, we can construct the effective curvature gravity theory from F-theory~\cite{bng,anom2}. In this section, we review construction of this effective curvature gravity from the F-theory. And in subsection 4.1, we consider the generalization for the case of the curvature 
 gravity on $AdS_5\times X_5$ with general $\Delta$.

In order to obtain the higher derivative terms in the action of F-theory compactifications, we use a duality chain
\begin{equation}
{\rm Heterotic\;}SO(32)/T^{2} \rightarrow {\rm Type\; I}/T^{2} \rightarrow {\rm Type IIB}/T^{2}/\mathbb{Z}_{2}, \label{dual7}
\end{equation}
where we break the gauge symmetry of $SO(32)$ into $SO(8)^{4}$ in Heterotic string theory and Type I string theory by turning on Wilson lines. 

In the ten dimensional supergravity action of Heterotic string theory, the terms which include Riemann tensors up to the quadratic order is 
\begin{equation}
S_{het}=\frac{1}{(2 \pi)^{7}l_{s}^{8}} \int dx^{10}\sqrt{g^{h}}e^{-2\phi^{h}}\left(R+\frac{l_{s}^{2}}{4}R_{MNLP}R^{MNLP}\right),
\label{eq:action_het}
\end{equation}
where $M,N,L,P=1,...,10$.
When one moves to Type I string theory, the duality relations are 
\begin{equation}
\phi^{h} = -\phi^{I},\;\;
g_{MN}^{h} = e^{-\phi_{I}} g_{MN}^{I}.
\end{equation}
Then, the action \eqref{eq:action_het} becomes
\begin{equation}
S_{I}=\frac{1}{(2 \pi)^{7}l_{s}^{8}} \int dx^{10}\sqrt{g^{I}}\left(e^{-2\phi_{I}} R + \frac{l_{s}^{2}}{4}e^{-\phi_{I}}R_{MNLP}R^{MNLP} \right),
\label{eq:action_typeI}
\end{equation}
and we have 32 D9-branes and 1 O9-plane. We also consider $N$ coincident 5-branes wrapped on the torus $T^{2}$. From such a configuration, one obtains $USp(2N)$ gauge symmetry with a hypermultiplet in the anti-symmetric representation on 5-branes. Besides we have 32 additional half-hypermultiplets in the {\bf 2N} representation of $USp(2N)$. 

As for the second step, we take two T-duality along the torus. Then, we have 16 D7-branes and 4 O-planes which are separated into four groups. The difference of the number of D7-brane comes from the fact that the charges on both sides are related as $\mu_{7}^{I}=\frac{1}{2}\mu_{7}^{IIB}$. Each group locate at the four fixed points of $T^{2}/\mathbb{Z}_{2}$. We consider the case where $T^{2}$ is large and focus on one fixed point. Other fields related to other fixed points are very massive and can be integrated out in the low energy field theory. Furthermore, $N$ 5-branes become $N$ 3-branes after the two T-dualities along $T^{2}$. Hence, the low energy field theory is described by $\mathcal{N}=2$ $USp(2N)$ gauge theory with four fundamental hypermultiplets and an additional hypermultiplet in the anti-symmetric representation. The duality relations between Type I and Type IIB string theory are 
\begin{equation}
V_{I}e^{-2\phi^{I}} = V_{IIB} e^{-2\phi^{IIB}},\;\; R_{I}^2 = \frac{l_{s}^{4}}{L^2},
\end{equation}
where $L$ is the radius of the torus $T^2$ in Type IIB theory and $V_I$ and $V_{II}$ are the volume of the torus in Type I and Type IIB theory, respectively. 
The action of Type IIB string theory becomes 
\begin{equation}
S_{IIB}=\frac{1}{(2 \pi)^{7}l_{s}^{8}} \int dx^{10}\sqrt{g^{IIB}}e^{-2\phi_{IIB}} R + \frac{1}{(2 \pi)^{7}l_{s}^{8}} \frac{(2\pi)^{2}l_{s}^{2}}{2}\int d^{8}x \sqrt{g^{IIB}} \frac{l_{s}^{2}}{4}e^{-\phi_{IIB}}R_{MNLP}R^{MNLP}.
\label{eq:action_typeIIB}
\end{equation}
Note that the second term in \eqref{eq:action_typeIIB} is proportional to $e^{-\phi^{IIB}}$ and the world volume is eight dimension. Hence, this term originates from the D7-brane and O7-plane. We also take into account the difference between the charge of D9-branes in Type I string theory and the charge of D9-branes in Type IIB string theory $\mu_{9}^{I} = \frac{1}{\sqrt{2}}\mu_{9}^{IIB}$. Since we will focus on the one of the four fixed points in $T^{2}/\mathbb{Z}_{2}$, we further multiply a factor $\frac{6}{24}=\frac{1}{4}$ to the second term in \eqref{eq:action_typeIIB}
\be
S_{IIB}=\frac{1}{(2 \pi)^{7}l_{s}^{8}} \int dx^{10}\sqrt{g^{IIB}}e^{-2\phi_{IIB}} R + \frac{1}{(2 \pi)^{7}l_{s}^{4}} \frac{(2\pi)^{2}}{2} \frac{1}{4} \frac{1}{4}\int d^{8}x \sqrt{g^{IIB}}e^{-\phi_{IIB}}R_{MNLP}R^{MNLP}.
\label{s1}
\ee

By the dimensional reduction of the 5-dimensional and 3-dimensional internal directions, \eqref{s1} reduces following effective five dimensional action plus cosmological constant $-12/\tilde{L}^2$ which can have the $AdS_5$ vacua as the solution:
\ba
&S_{IIB}=\frac{L^5vol(X_{\Delta=2})e^{-2\phi_{IIB}}}{(2 \pi)^{7}l_{s}^{8}} \int dx^{5}\sqrt{g^{IIB}} \Big(R+\dfrac{12}{{\tilde{L}}^{2}}\Big) \\ \notag
&+ \frac{L^3vol(S^{3})}{(2 \pi)^{7}l_{s}^{4}} \frac{(2\pi)^{2}}{2} \frac{1}{4} \frac{1}{4}\int d^{5}x \sqrt{g^{IIB}}e^{-\phi_{IIB}}R_{MNLP}R^{MNLP},
\label{s3}
\ea
Note that the higher derivative terms also correct the cosmological constant.
As pointed out in~\cite{bng,Fukuma:2001uf,Fukuma:2002sb}, 
 the cosmological constant $-12/\tilde{L}^2$, which is different from $-12/L^2$ via $1/N$ correction, is determined later for the gravity dual to $\mathcal{N}=2$ $USp(2N)$ gauge theory and also depends on the field redefinition. 

Note that $\frac{(2\pi)^{2}}{2}=vol(T^{2}/\mathbb{Z}_{2})/L^{2}$ and we can formally write the second term with a coefficient $vol(X_{5,\Delta=2})$. Hence this factor can be factored out and the action \eqref{s3} becomes
\be
S_{IIB}=\frac{L^5vol(X_{5,\Delta=2})e^{-2\phi_{IIB}}}{(2 \pi)^{7}l_{s}^{8}} \int dx^{5}\sqrt{g^{IIB}}\left( R+\dfrac{12}{\tilde{L}^2} +  \frac{L^2}{16N}R_{MNLP}R^{MNLP} \right), 
\label{eq:action_typeIIB_4}
\ee
where we used the relation $L=(4\pi g_s\Delta N)^{1/4}l_s$ with $\Delta =2$.

Thus, the coefficient $a_3=L^2/(16N)$ for the gravity dual on $AdS_5\times S^5/\mathbb{Z}_2$ is determined using the duality chain \eqref{dual7}. 

\subsection{Power counting}

So far, we have only focused on the two terms, namely the Einstein-Hilbert action in the bulk and the quadratic term of the Riemann tensor on 7-branes. Let us see which terms will contribute to the $A$-type anomaly from the entanglement entropy. For simplicity, we call $R$ as either $R_{MNKL}, R_{MN}, R$. Since the $A$-type anomaly depends on the parameters $N, \Delta$ and $n_{7}$, we will focus on the $N, \Delta$ and $n_{7}$ dependence in the action. In general, we have the terms
\be
\frac{e^{-2\phi_{IIB}}}{l_{s}^{8}} \int d^{10}x \; \sqrt{g^{IIB}} l_{s}^{2(k-1)} R^{k} + \frac{n_{7}e^{-\phi_{IIB}}}{l_{s}^{8}} \int d^{8}x \; \sqrt{g^{IIB}} l_{s}^{2k} R^{k}.
\label{eq:action_counting1}
\ee
The first term of \eqref{eq:action_counting1} is the one in the bulk action and the second term of \eqref{eq:action_counting1} is the one in the 7-brane action. Hence, the second term is proportional to the number of 7-branes. Remember that supersymmetry protects string loop corrections up to the lowest order of the DBI terms and then the second term only depending on 
 $e^{-\phi_{IIB}}$ arises from 7-branes~\cite{Tseytlin:1999dj}. By the dimensional reduction on $X_{5}$, \eqref{eq:action_counting1} becomes
\be
\frac{e^{-2\phi_{IIB}}L^{5}}{l_{s}^{10-2k} \Delta} \int d^{5}x \; \sqrt{g^{IIB}} R^{k} + \frac{e^{-\phi_{IIB}}n_{7} L^{3}}{l_{s}^{8-2k}} \int d^{5}x \; \sqrt{g^{IIB}}R^{k}.
\ee
The factor $\frac{1}{\Delta}$ appears from the dimensional reduction of $vol(X_{5}) \propto \frac{1}{\Delta}$. By applying the Wald formula \eqref{CEN5} for the entanglement entropy and the $A$-type anomaly, the $A$-type anomaly is roughly expressed as
\begin{equation}
a \sim \frac{e^{-2\phi_{IIB}}L^{10-2k}}{l_{s}^{10-2k} \Delta} + \frac{e^{-\phi_{IIB}}n_{7} L^{8-2k}}{l_{s}^{8-2k}}.
\label{eq:action_counting3}
\end{equation}
The first term originates from the term in the bulk action and the second term originates from the term in the 7-brane action. Because of the relation $L \propto N^{\frac{1}{4}}$, $\mathcal{O}(N^{2})$ contribution only comes from the $k=1$ term in the bulk action, namely the Einstein-Hilbert term. The $\mathcal{O}(N)$ can come from the $k=3$ terms in the bulk action and $k=2$ term in the 7-brane action. However, there are no $k=3$ terms in the five dimensional action since higher derivative correction for the Type IIB theory starts with $\mathcal{O}(R^4)$~\cite{Gris,Ogawa:2011fw}. Here, $k=3$ term may be generated in terms of the dimensional reduction of $k=4$ term. However, it has the role of $k=4$ term since the dimensional reduction does not change the result. In the dual SCFT side, moreover, the coefficient of $k=3$ term corresponding to the three point function of the stress tensor, which would vanish for any SCFTs~\cite{Hofman:2008ar}.
 Hence $\mathcal{O}(N)$ contribution only comes from the $k=2$ term in the 7-brane action. Therefore, the first term in \eqref{eq:action_typeIIB} will generate a $\mathcal{O}(N^{2})$ contribution to the $A$-type anomaly and the second term in \eqref{eq:action_typeIIB} will make a $\mathcal{O}(N)$ contribution. From the general expression of \eqref{eq:action_counting3} and concentrating only on the $\mathcal(N^{2})$ term and the $\mathcal{O}(N)$ in the $A$-type anomaly, we can also determine the $N, \Delta$ and $n_{7}$ dependence of the $A$-type anomaly. The $A$-type anomaly should behave as 
\begin{equation}
a = \alpha_{1}N^{2}\Delta + \alpha_{2} N \Delta n_{7},
\label{eq:a-anomaly_general}
\end{equation}
where $\alpha_{1}$ and $\alpha_{2}$ are numerical coefficients. 

Finally, it is interesting to investigate the $1/N^2$ correction. Since the bulk $R^3$ term is not allowed for any SCFTs and since the more higher derivative terms seem not to contribute the central charges, the $1/N^2$ correction should be included in the gravity theory with curvature squared terms. In other words, this correction can be included not in more higher derivative terms but in coefficients of the gravity theory such as the cosmological constant.

\subsection{Determination of $\tilde{L}^2$}
Next, we determine the cosmological constant $-12/\tilde{L}^2$ by comparing the central charges $a,c$ in SCFT side with that in the holographic Weyl anomaly~\cite{bng,ode,holo4}. This method 
 includes the review of~\cite{deBoer:2011wk,Hung:2011xb} up to the scaling transformation of the coefficients of the gravity theory.
In the field theory side, we can identify $a,c$-charges from the coefficients of the trace anomaly in 4-dimension as follows:
\ba
\la T^{a}{}_{a}\lb =\dfrac{c}{16\pi^2}(R_{abcd}R^{abcd}-2R_{ab}R^{ab}+R^2/3)-\dfrac{a}{16\pi^2}(R_{abcd}R^{abcd}-4R_{ab}R^{ab}+R^2),
\ea
where $a,b=0,..,3$ and $R_{abcd}$ is the 4-dimensional Riemann tensor. This coefficients $a$ and $c$ can be derived using holographic Weyl anomaly formula. We start with the curvature gravity \eqref{ACT6}.
We first derive the AdS vacua for \eqref{ACT6} by solving the EOM.
In the EOM for the action \eqref{ACT6}, the terms including the covariant derivatives vanish in the critical points describing $AdS_5$. Leaving $a_1$, $a_2$, and $a_3$,
 the EOM becomes
\ba
&-\dfrac{1}{2}g_{\mu\nu}\Big(a_1R^2+a_2R_{\alpha\beta}R^{\alpha\beta}+a_3R_{\alpha\beta\gamma\delta}R^{\alpha\beta\gamma\delta}+R+\dfrac{12}{\tilde{L}^2}\Big) \nonumber \\
&+2a_1RR_{\mu\nu}+2a_2R_{\mu\alpha}R_{\nu}{}^{\alpha}+2cR_{\mu\alpha\beta\gamma}R_{\nu}{}^{\alpha\beta\gamma}+R_{\mu\nu}=0. \label{EOM11}
\ea
Substituting the formula in the appendix A, 
 \eqref{EOM11} is rewritten as
\ba
&1 l^4-12 \tilde{L}^2l^2+(80a_1+16a_2+8a_3)\tilde{L}^2=0. \label{EQ26}
\ea
Substituting $a_3=L^2/16N$, $a_2=a_1=0$ into above the equation, \eqref{EQ26} is solved by
\ba
&l^2=\tilde{L}^2\dfrac{1+\sqrt{1-\dfrac{4}{3\tilde{L}^2}(20a_1+4a_2+2a_3)}}{2}, \\
&=\dfrac{\tilde{L}^2}{2}\Big({1+\sqrt{1-\dfrac{L^2}{6N\tilde{L}^2}}}\Big).
\label{CUR12}
\ea
where $l$ is the AdS radius in an AdS vacua obtained from \eqref{EOM11}.

According to~\cite{ode} and using \eqref{CUR12}, $a,c$-charges are given by
 \begin{align}
&a=\dfrac{\pi^2 l^3}{\kappa_5^2}(1-4a_3/l^2), \label{ACA26} \\
& c=\dfrac{\pi^2 l^3}{\kappa^2_5}(1+4a_3/l^2). \nonumber
\end{align}
It can be shown that these central charges are invariant under the scaling transformation of the metric $g_{\mu\nu}\to e^{2a}g_{\mu\nu}$.
Under this scaling transformation, the coefficients in the action \eqref{ACT6} are transformed as $\kappa_5^2\to \kappa_5^2e^{-3a}$, $\tilde{L}\to \tilde{L}e^{-a}$, $l\to  le^{-a}$, and $a_i\to a_ie^{-2a}$. Thus, central charges $a,c$ are invariant under above scaling transformation. For later purpose, we leave this dependence of $e^a$ in the coefficient of parameters.

To recover the central charges in the field theory side~\cite{at} with $\Delta=2$ 
\begin{align}
&a=\dfrac{\Delta N^2}{4}+\dfrac{(\Delta -1)N}{2}-\dfrac{1}{24},\quad c=\dfrac{\Delta N^2}{4}+\dfrac{(\Delta -1)3N}{4}-\dfrac{1}{12}, \label{CEN47}
\end{align}
we should solve the simultaneous equations~\eqref{ACA26}.
 $l$ and $a_3$ are then given by
\ba
&l^2=L^2e^{-2a}\Big(1+\dfrac{5}{4N}-\dfrac{1}{8N^2}\Big)^{{2}/{3}}, \label{LLL40} \\
&a_3=\dfrac{L^3e^{-2a}}{16l}\Big(\dfrac{1}{N}-\dfrac{1}{6 N^2} \Big).
\ea
The radius $\tilde{L}$ is determined by solving \eqref{CUR12} and including the factor $e^{-a}$ as follows:
\ba
\tilde{L}^2=\dfrac{4l^4N(144Nl^2-6L^2)e^{-2a}}{(L^2-24Nl^2)^2},
\ea 
where we used $l$ in \eqref{LLL40}.
To recover $a_3=L^2/16N$ obtained from F-theory, moreover, we should set
\ba
e^{-2a}=\dfrac{l}{L\Big(1-\dfrac{1}{6N}\Big)}.
\ea
It is concluded that the Taylor-series of $\tilde{L}$ and $l$ in terms of $1/N$ are given by  
\ba
l^2=  {L}^{2}+\dfrac{17L^2}{12N}+\dfrac{L^2}{9N^2}+O \left( {N}^{-3} \right),
\ea
\ba
\tilde{L}^2=  {L}^{2}+\dfrac{35L^2}{24N}+\dfrac{79L^2}{576N^2}+O \left( {N}^{-3} \right). 
\ea

\subsection{General cases}
For general cases with the gravity dual on $AdS_5\times X_5$, moreover, the coefficient $a_3$ should be determined using the normalization of $a_3=L^2/(16N)$ for $\Delta =2$ since the origin of this squared curvature term is 7-brane in the Type IIB theory~\cite{at} as follows: 
\be
a_3=\dfrac{(1-1/\Delta)L^2}{8N}.
\ee 
This coefficient has the information of the number of 7-branes $n_7=12(1-1/\Delta)$.
This formula is also obtained in~\cite{Buchel:2008vz} requiring that the difference of two central charges $a-c$ in the field theory side is reproduced using the effective curvature gravity. 

 For the gravity dual on $AdS_5\times X_5$, thus, we should substitute $a_1=0$, $a_2=0$, and $a_3=(1-1/\Delta)L^2/8N$ into \eqref{ACT6} and the 
 effective curvature gravity action becomes
\ba
S=\dfrac{1}{2\kappa_{5}^2}\int d^5x\sqrt{-g}\Big(R+\dfrac{12}{\tilde{L}^2}+\dfrac{(1-1/\Delta)L^2}{8N}R_{\mu\nu\rho\lambda}R^{\mu\nu\rho\lambda} \Big),
\label{ACT7} \ea
where $\tilde{L}$ can be determined by using the method in previous section and by using \eqref{CEN47}. In terms of the $1/N$ expansion, the $AdS$ radius $l$ and $\tilde{L}$ are given by   
\ba
l^2=  {L}^{2}+{\frac {{L}^{2} \left( 15{\Delta}^{
2}-29\Delta+15 \right) }{ 6N\left( \Delta-1 \right) \Delta}}+{
\frac {{L}^{2} \left( 6{\Delta}^{2}-11\Delta+6 \right) }{
 36N^2\left( \Delta-1 \right) ^{2}\Delta}}+O \left( {N}^{-3} \right),
\ea
\ba
\tilde{L}^2=  {L}^{2}+{\frac {{L}^{2} \left( 31{\Delta}^
{2}-60\Delta+31 \right) }{12N \left( \Delta-1 \right) \Delta}}+{\frac {{L}^{2} \left( 11{\Delta}^{4}-18{\Delta}^{3}+18
{\Delta}^{2}-18\Delta+11 \right) }{144N^2 \left( \Delta-1
 \right) ^{2}{\Delta}^{2}}}+O \left( {N}^{-3} \right). 
\ea

\section{Computation of the $A$-type anomaly from the entanglement entropy} 

In this section, we compute the holographic EE for $d=4$ $\mathcal{N}=2$ SCFT with its gravity dual using holographic EE formula.
   
 We first consider $O(N^2)$ part of the anomaly. Since the curvature correction and $1/N$ correction do not contribute the central charge of this order,  we can ignore the $1/N$ corrections for $\tilde{L}$, $\kappa_5$, $a_3$ and the curvature corrections ($a_3=0$). Thus, central charges $a$ equals $c$ since we consider only Einstein-Hilbert term \eqref{AC09}.

An useful metric of $AdS_5$ is a hyperbolic foliation of $AdS_5$ and is given by \eqref{MET05}.
 The horizon entropy can be calculated using Wald's entropy
\ba
S=-2\pi \int_{horizon}d^{d-1}x\sqrt{h}\dfrac{\partial \mathcal{L}}{\partial R^{\mu\nu}{}_{\rho\lambda}}\epsilon^{\mu\nu}\epsilon_{\rho\lambda},
\ea
where $\epsilon_{ab}$ denotes the binormal to the horizon. Using the wedge product, the binormal is represented by $d\tau\wedge d\rho$.
 The integrand for the action \eqref{AC09} can be rewritten as
\ba
\dfrac{\partial \mathcal{L}}{\partial R^{\mu\nu}{}_{\rho\lambda}}\epsilon^{\mu\nu}\epsilon_{\rho\lambda}\Big|_{AdS}=-\dfrac{vol(X_5)}{\kappa_{10}^2}L^5,
\ea
where we substituted the only non-zero component of the binormal $\epsilon_{t\rho}=1$ or the relation $\epsilon_{\mu\nu}\epsilon^{\mu\nu}=-2$. 
Thus, using the relation proven in~\cite{Myers:2010xs,Myers:2010tj}
\ba
\dfrac{\partial \mathcal{L}}{\partial R^{\mu\nu}{}_{\rho\lambda}}\epsilon^{\mu\nu}\epsilon_{\rho\lambda}\Big|_{AdS}=-\dfrac{a_4}{\pi^2 L^3}, \label{ACA8} \ea
the central charge $a_4^{(2)}$, namely the $O(N^2)$ contribution for the $A$-type anomaly is computed as follows:
\ba
a_4^{(2)}=\dfrac{\pi^2L^8vol(X_5)}{\kappa_{10}^2}=\dfrac{\Delta N^2}{4}. \label{CEN32}
\ea
This central charge correctly reproduces the $A$-anomaly of dual $\mathcal{N}=2$ SCFT. Remember that \eqref{CEN32} agrees with the central charges $a_4$ using the GKP-W relation~\cite{anom}.

\subsection{$1/N$ correction of $A$-type anomaly}

 For the action \eqref{ACT7}, the horizon entropy becomes
\begin{align}
&S^E=\dfrac{2\pi}{\kappa_5^2}\Big(1-4a_3  \dfrac{1}{r_h^2}\Big)\oint d^{d-1}x\sqrt{h(r_h)} \nonumber \\
&=\dfrac{2\pi}{\kappa_5^2}\Big(1-\dfrac{(1-1/\Delta)L^2}{2Nl^2} \Big)\oint d^{3}x\sqrt{h(r_h)}, \label{ENT22}
\end{align}
where in the last equality, we used $l=r_h$.

According to~\eqref{ACA8}, the $A$-type anomaly is computed as follows:
\ba
a_4=\dfrac{\pi^2{l}^3}{\kappa_5^2}\Big(1-\dfrac{(1-1/\Delta)L^2}{2Nl^2}\Big)=\dfrac{\Delta N^2}{4}+\dfrac{(\Delta-1) N}{2}-\dfrac{1}{24}, \label{ATY36}
\ea
where we used \eqref{DEF10} and \eqref{CEN32}. The central charge \eqref{ATY36} is the same as the holographic Weyl anomaly \eqref{ACA26}.  It is concluded that central charge \eqref{ATY36} agrees with the $A$-type anomaly in the CFT side.

Then, we consider the field redefinition of the curvature theory \eqref{ACT7} to avoid the graviton ghost in the subleading order of $l_s^2/L^2$ expansion.
As pointed out in~\cite{Myers:2010tj}, the linearized gravition equation of \eqref{ACT7} is not second order but is 4-th order. So, there appear gravition ghosts for 4th order equations, which lead to the non-unitary operator in the dual SCFTs. 
 Using the field redefinition~\cite{Buchel:2008vz,kp}, however, we show that in the subleading order of $l_s^2/L^2$ expansion, we can transform the curvature gravity \eqref{ACT7} to the Gauss-Bonnet gravity, which has second order linearized graviton equations and preserves unitarity. 
We transform the metric as
\ba
g_{\mu\nu}\to g_{\mu\nu}+b_1g_{\mu\nu}R+b_2R_{\mu\nu},
\ea
where $b_1=-(1-1/\Delta)L^2/12N$ and $b_2=(1-1/\Delta)L^2/2N$.
Then, the gravity action \eqref{ACT7} can be transformed to
\ba
S=\dfrac{1}{2\kappa_{5}^2\beta}\int d^5x\sqrt{-g}\Big(R+\dfrac{12\beta}{ \tilde{L}^2}+\dfrac{(1-1/\Delta)L^2}{8N}(R_{\mu\nu\rho\lambda}R^{\mu\nu\rho\lambda}+R^2-4R_{\mu\nu}R^{\mu\nu})+... \Big),
\label{ACT19} \ea 
where $\beta^{ -1}=1+(1-1/\Delta)/(2N)$ and dots describe the terms of $\mathcal{O}(R^3)$ and terms of $O(N^{-2})$ corrections.

Using \eqref{EOM11}, the $AdS$ radius of the $AdS$ solution obtained from \eqref{ACT19} is given by
\ba
&\tilde{l}^2=\tilde{L}^2\dfrac{1+\sqrt{1-\dfrac{4\beta^2}{3\tilde{L}^2}(20a_1+4a_2+2a_3)}}{2\beta}, \\
&=\dfrac{\tilde{L}^2}{2\beta}\Big({1+\sqrt{1-\dfrac{1-1/\Delta}{N}}}+O(N^{-2})\Big).
\label{CUR2}
\ea

 For the action \eqref{ACT19}, the horizon entropy\footnote{In the context of Kerr/CFT correspondence~\cite{Guica:2008mu}, the Wald's entropy of a $d=5$ rotating black hole in the presence of Gauss-Bonnet terms coincides with the Cardy's formula~\cite{Hayashi:2011uf}.} becomes
\begin{align}
&S^E=\dfrac{2\pi}{\kappa_5^2\beta}\Big(1-12a_3\beta  \dfrac{1}{r_h^2}\Big)\oint d^{d-1}x\sqrt{h(r_h)} \nonumber \\
&=\dfrac{2\pi}{\kappa_5^2\beta}\Big(1-\dfrac{3(1-1/\Delta)\beta}{2N} \Big)\oint d^{3}x\sqrt{h(r_h)}, \label{ENT22x}
\end{align}
where in the last equality, we used ${L}=r_h$ of the leading order of $N$.

According to~\eqref{ACA8}, the $A$-type anomaly is computed as follows:
\ba
a_4=\dfrac{\pi^2\tilde{l}^3}{\kappa_5^2\beta}\Big(1-\dfrac{3(1-1/\Delta)\beta}{2N}\Big)=\dfrac{\Delta N^2}{4}+\dfrac{(\Delta-1) N}{2}+O(1), \label{ATY36x}
\ea
where we used \eqref{DEF10} and \eqref{CEN32}. Since \eqref{ATY36x} is the same form as \eqref{ACA26}, this central charge is invariant under the scaling transformation $g_{\mu\nu}\to e^{2a}g_{\mu\nu}$.
Central charge \eqref{ATY36x} agrees with the $A$-type anomaly in the CFT side up to order $N$~\eqref{CEN47}.

Lastly, we discuss the holographic c-theorem proposed in~\cite{Myers:2010tj,Liu:2011ii}, where the usual $c$-theorem in quantum field theories~\cite{Zam,Cardy:1988cwa,Osborn:1989td,Jack:1990eb,Komargodski:2011vj} states that central charges decrease in the IR. To describe the holographic $c$-theorem, the bulk matter should satisfy two conditions, namely, the null energy condition $-(T^t_t-T^r_r)\ge 0$ and $\sigma'\ge 0$ where
\ba
\sigma =\dfrac{16a_1+5a_2+4a_3}{16}R', \label{NUL21}
\ea
where $R'$ is the derivative of the Ricci scalar along the AdS radial direction. Here, we used the convention of the metric in~\cite{Liu:2011ii}. 
 The Gauss-Bonnet gravity \eqref{ACT19} clearly satisfies the second condition. 
So, it is concluded that if we set the coefficients of the curvature square terms properly using the field-redefinition, we may only consider the bulk null energy condition $-(T^t_t-T^r_r)\ge 0$ to be consistent with the holographic c-theorem.

\section{Discussion}
In this paper, we computed the holographic EE in the effective curvature gravity dual to the $\mathcal{N}=2$ SCFTs in the F-theory using the Wald's entropy~\cite{Myers:2010tj} and confirmed that it is consistent with the proposal in~\cite{Myers:2010tj}. We realized the $A$-type anomaly including the $1/N^2$ corrections in the SCFT side from the log term of the holographic EE not using the Gauss-Bonnet type but using the general curvature squared gravity, where it was also discussed in~\cite{deBoer:2011wk,Hung:2011xb}. Here, the curvature term of this theory can be transformed into the Gauss-Bonnet term \eqref{ACT19} in the subleading order of $l_s^2/L^2$ expansion. Our new analysis in section 4.1 also shows that
 the $1/N^2$ correction is included not in the higher derivative terms of $O(R^3)$ but in the gravity theory with general curvature squared terms. While the $A$-type anomaly can be computed by using the $a$-maximalization in supersymmetric theories, the analysis of this paper gives the interesting method to compute the entanglement entropy and the $A$-type anomaly. 
 Precisely, we chose the cosmological constant $-12/\tilde{L}^2$ to reproduce the Weyl anomaly in the dual SCFTs using the holography since  this cosmological constant has not been determined from the 10-dimensional theory yet. 

In appendix B, we considered the $1/N$ correction from the higher derivative term of the probe $D_4$-type 7-brane, where we can take the $g_s\to 0$ limit. This $1/N$ correction in our approximation agreed with the result in the CFT side up to a factor 15/16. We leave the similar analysis for more complicated cases $G=E_6,E_7,..$ with a constant coupling.

To obtain consistent results, we also discussed a holographic c-theorem~\cite{Myers:2010tj,Liu:2011ii}.  Using the field redefinition properly, we showed that the Gauss-Bonnet gravity \eqref{ACT19} satisfies the null energy condition in the subleading order of the $l_s^2/L^2$ expansion. This means that using the field redefinition, a theory with non-trivial $\sigma$ \eqref{ACT7} may be equal to the Gauss-Bonnet gravity with $\sigma=0$ \eqref{ACT19} of the finite order of $N$.\footnote{ Note that as another direction, the action \eqref{ACT19} is the Gauss-Bonnet action and we can apply the holographic EE on the $AdS_5$ soliton~\cite{Ogawa:2011fw} using the action \eqref{ACT19}. We then notice an instability of the holographic EE for $a_3>0$.  However, a mechanism of the supersymmetry seems to work for our theory for $\mathcal{N}=2$ SCFTs and we do not mind this instability.  } 
 See also~\cite{Gubser:2002zh,Fujita:2008rs} for the paper discussing the holographic c-theorem in the context of double trace deformation.

\bigskip
\noindent {\bf Acknowledgments:}  We would like to thank A. Buchel, C. Hoyos, M. Kaminski, A. Karch, T. Nishioka, N. Ogawa, H. Ooguri, S. Sugimoto, Y. Tachikawa and T. Takayanagi for discussions and comments. I would like to thank H. Hayashi and T. S. Tai for collaboration in the initial stage of this project and for helpful discussions.  MF is supported by the postdoctoral fellowship program of the Japan Society for the Promotion of Science
(JSPS), and partly by JSPS Grant-in-Aid for JSPS Fellows No. 22-1028.

\appendix

\section{$AdS_{d+1}$ spacetime}
In this appendix, we review the Riemann tensor, Ricci tensor, and Ricci scalar of the maximally symmetric anti-de Sitter space.
The Einstein-Hilbert Lagrangian is
\be
\mathcal{L}=\sqrt{-g}(R-2\Lambda).
\ee
The solutions we are looking for are called maximally symmetric 
 and satisfy the conditions
 \begin{align}
 &R_{\lambda\mu\kappa\nu}=-\dfrac{1}{L^2}(g_{\lambda\kappa}g_{\mu\nu}-g_{\lambda\nu}g_{\kappa\mu}),\ R_{\mu\nu}=-\dfrac{d}{L^2}g_{\mu\nu}, \\
 &R=-\dfrac{d(d+1)}{L^2},
 \end{align}
where $L^2=-d(d-1)/2\Lambda$. Remind that $\Lambda <0$ for the anti-de Sitter space.

\section{Correction from 7-branes}
To compute $O(N)$ contributions for the $A$-type anomaly, we can consider the higher derivative corrections on the 7-branes in $AdS_5\times X_{5}$ for $\Delta =2$ instead of the 5-dimensional effective curvature action. For this $D_4$-type 7-brane, we can take the limit $g_s\to 0$ to suppress the quantum correction while it keeps the 't Hooft coupling $g_sN$ fixed. In the present case,  there are no backreactions of the 7-brane because of the tadpole cancellation between D7-branes and O7-brane. In this section, we show that the higher derivative term on the 7-brane at the singularity can reproduce the $O(N)$ correction for the $A$-type anomaly of the SCFT dual to $AdS_5\times S^5/\mathbb{Z}_2$  up to a factor of 15/16. 

Relevant curvature corrections~\cite{sign} are given by
\ba
&S=-n_7\tau _7\int d^{8}\xi e^{-\phi}\sqrt{-G_{\alpha\beta}}(1-\dfrac{(4\pi^2 l_s^2)^2}{3\times 2^8\pi^2}(R_{\alpha\beta\gamma\delta}R^{\alpha\beta\gamma\delta}-2\hat{R}_{\alpha\beta}\hat{R}^{\alpha\beta}-R_{ab\alpha\beta}R^{ab\alpha\beta}+2\hat{R}_{ab}\hat{R}^{ab}) \nonumber \\
&+O(l_s^4)),
\ea
where $\alpha,\beta$ are the tangent space indices and $a,b$ are the normal space indices. Here, 
$\hat{R}_{\alpha\beta}$, $\hat{R}_{ab}$ and $\hat{R}$ are obtained only by contracting tangent indices. The above formula is only valid for geodesic embeddings of the world-volume and is enough for our case. 

After the dimensional reduction for $S^3$, the first two terms of the higher curvature terms contribute to the action.\footnote{Here, we ignored the mixed components of the curvature term.} The coefficient of $R_{\mu\nu\rho\lambda}R^{\mu\nu\rho\lambda}$ becomes 
\ba
n_7\tau_7 L^3 Vol(S^3) \dfrac{(4\pi^2 l_s^2)^2 }{3\cdot 2^8\pi^2 g_s}=\dfrac{N}{64\pi^2 L},
\ea
where we used the equation $n_7=6$. 
And the reduced action couples with the gravity action \eqref{AC09} as follows:
\ba
S=\dfrac{1}{2\kappa_{5}^2}\int d^5x\sqrt{-g}\Big(R+\dfrac{12}{ L^2}+\dfrac{L^2}{16N}\Big(R_{\mu\nu\rho\lambda}R^{\mu\nu\rho\lambda}-2R_{\mu\nu}R^{\mu\nu}\Big) \Big),
\label{ACT27} \ea
Remember that this coefficient $a_3$ also agrees with the relative coefficient used in~\cite{bng,anom2,at}. To solve the EOM, we should include the contribution for the higher derivative correction on  the 7-branes to include the subleading corrections. And we obtain the different AdS vacua with the different AdS radius $l$ since this term is the subleading order in terms of $N$. 

It will be interesting to perform the field-redefinition of above the gravity as follows:
\ba
g_{\mu\nu}\to g_{\mu\nu}+c_1R_{\mu\nu}-\dfrac{c_1}{3}g_{\mu\nu}R,
\ea
where $c_1=-L^2/8N$. Then, we obtain the following gravity theory up to the subleading order of $1/N$:
\ba
S=\dfrac{1}{2\kappa_{5}^2\gamma}\int d^5x\sqrt{-g}\Big(R+\dfrac{12\gamma}{ L^2}+\dfrac{L^2}{16N}R_{\mu\nu\rho\lambda}R^{\mu\nu\rho\lambda}... \Big),
\ea
where $\gamma=(1+1/2N)^{-1}$. Note that this gravity theory has the same form as the effective curvature gravity \eqref{ACT7} up to the field redefinition.

Substituting $a_3=L^2/16N$, $a_2=0$, $a_1=0$ into the Ward's formula \eqref{ENT22}, the contribution of the anomalies from this worldvolume action becomes
\ba
a=\dfrac{ N^2}{2}+\dfrac{15N}{32}+O(1).
\ea
Thus, the $O(N)$ contribution agrees with the field theory result by the factor 15/16.

\end{document}